\input epsf

\def\eps@scaling{.95}
\def\epsscale#1{\gdef\eps@scaling{#1}}

\def\plotoneb#1{\centering \leavevmode
\epsfysize=7in \epsfbox{#1}}

\def \myplotone#1 {\plotone{#1}}
\def \myplotoneb#1 {\plotoneb{#1}}
\def \myplotfiddle#1 {\plotfiddle{#1}{512pt}{0}{100}{100}{-288}{-140}}

\def \grad {{\vec \nabla}}
\def \laplacian {{\nabla^2}}
\def \etal {{\it et al. }}
\def \lhood {{\cal L}}

\documentstyle[aaspp4,12pt]{article}

\slugcomment{CfPA 95-th-35/Submitted to Ap.J.}

\begin{document}

\title{Unbiased Cluster Lens Reconstruction}
\author{Gordon Squires\altaffilmark{1}\altaffilmark{,2} and Nick
Kaiser\altaffilmark{3}}
\altaffiltext{1}{Department of Physics, University of Toronto,
60 St.\ George St., Toronto, Ontario, M5S 1A7}
\altaffiltext{2}{Center for Particle
Astrophysics, University of California, Berkeley, CA 94720\\
{\tt squires@magicbean.berkeley.edu}}
\altaffiltext{3}{Canadian Institute for Advanced Research and
Canadian Institute for Theoretical Astrophysics, University of Toronto,
60 St.\ George St., Toronto, Ontario, M5S 1A7\\
{\tt kaiser@cita.utoronto.ca}}

\begin{abstract}
Weak lensing observations measure the shear field $\gamma_\alpha$,
and hence the gradient of the dimensionless surface density $\kappa$.
We present several new algorithms to recover $\kappa$
{}from shear estimates on a finite region and compare how they perform
with realistically noisy data. The reconstruction methods
studied here can be divided into two classes:
direct reconstruction and regularized inversion techniques.
The direct reconstruction techniques express
the surface density as a two-dimensional integral of the shear field:
$\kappa(\vec r_0) = \int d^2 r K_\alpha(\vec r; \vec r_0)
\gamma_\alpha(\vec r)$.
This allows one to construct an estimator for $\kappa$
as a discrete sum over background galaxy ellipticities which is straightforward
to implement, and allows a rigorous yet simple estimate of the noise
arising {}from random intrinsic background galaxy ellipticities.
We study three types of direct reconstruction methods: 1)
$\kappa$-estimators that measure the surface density at any given
target point relative to the mean value in some reference region 2) a
method that explicitly attempts to minimize the rotational part of
$\nabla \kappa$ that is due to noise and 3) a novel, exact Fourier-space
inverse gradient operator. We also develop two `regularized
maximum likelihood' methods, one of which employs the conventional
discrete Laplacian operator as a regularizer and the other uses
regularization of all components in Fourier space.
We compare the performance of all the estimators by means of simulations
and noise power analysis. A general feature of these unbiased
methods is an enhancement of the low-frequency noise power
which, for some of the methods, can be quite severe.  We find
the best performance is provided by the maximum likelihood method
with Fourier space regularization, although some of the other methods
perform almost as well.
\end{abstract}


\section{Introduction}

Clusters of galaxies, acting as gravitational lenses, introduce a statistical
anisotropy in the shapes of faint background galaxies.  In the weak
distortion regime it is possible to construct
(see e.g. \cite{ksb95}) a `polarization' statistic $e_\alpha$
--- a certain measure of the ellipticity of the background galaxy ---
whose expectation value is proportional to, and therefore provides a
direct measurement of, the gravitational shear
$\langle e_\alpha \rangle =
\gamma_\alpha \equiv \{(\phi_{,11} - \phi_{,22}) / 2, \phi_{,ij}\}$,
where $\phi$ is the surface potential for the lens.
The precision with which one can determine $\gamma_\alpha$ is of course
limited by the number of background galaxies, and care must be taken to allow
for systematic bias {}from seeing and removal of artificial anisotropy, but
several groups have now shown that the shear can be detected to
a reasonably high level of significance. The list of shear fields
mapped around clusters
includes A1689 (\cite{tyson90,tyson95}), A2218 (\cite{squires95}),
Cl1409+52 (\cite{tyson90}), MS1224+20 (\cite{fahlman94}),
Cl0024+17 (\cite{bonnet94a,mellier94}),
Cl1455+22 and Cl0016+16 (\cite{smail94,smail95}).

The lens surface potential is related to the dimensionless surface density
$\kappa \equiv \Sigma / \Sigma_{\rm crit}$ by $\laplacian \phi = 2 \kappa$.
In an earlier paper (Kaiser \& Squires 1993; hereafter KS93) we
proposed a method for reconstructing
the surface density {}from the shear.  Writing the shear as $\gamma_\alpha =
{\cal D}_\alpha \kappa$, where the 2-component integro-differential operator
is ${\cal D}_\alpha \equiv \{\partial_1^2 - \partial_2^2, 2 \partial_1
\partial_2 \} \nabla^{-2}$ one can readily show that
$\kappa = {\cal D}_\alpha \gamma_\alpha$.  The Green's function for this
operator is $-\chi_\alpha(\vec r) / \pi$ with $\chi_\alpha(\vec r) =
\{r_1^2 - r_2^2, 2r_1 r_2 \} / r^4$,
so we obtain
\begin{equation}	\label{eq:ks93}
\kappa(\vec r_0)  = -{1\over \pi} \int d^2 r \chi_\alpha(
\vec r - \vec r_0) \gamma_\alpha(\vec r)
\end{equation}
and this 2-dimensional
convolution integral can in turn be replaced by a discrete sum
over the background galaxies to give the estimator
\begin{equation}	\label{eq:ks93kappahat}
\hat \kappa(\vec r_0)  = - {1\over \overline n \pi} \sum_g
\chi_\alpha(\vec r_g - \vec r_0) e_\alpha
\end{equation}
where $\overline n$ is the mean surface number density of the background
galaxies.

Unfortunately the kernel $\chi_\alpha$ is infinite in extent, so this
requires, strictly speaking, data extending to infinity. With finite
data the estimate as written above becomes biased,
as discussed in KS93 and illustrated in the simulations presented there.
Near the center of
the field on which we have data, this simply results in a nearly
constant suppression in the reconstructed surface density, and since
the baseline surface density is ambiguous in any case this is relatively
benign.  However, near the edge of the data the {\sl shape\/}
becomes biased and one typically finds the density goes negative
in a trough around the cluster
and then rises again as one approaches the edge. The algorithm effectively
tries to generate a lens that has a shear pattern like that observed
within the region surveyed, but which falls to zero outside the data
boundary. One immediate consequence of this is that there must be
zero net mass in the reconstruction.  Because of this we have tended to
augment our 2-dimensional
reconstructions by aperture mass measurements (\cite{fahlman94,squires95})
which do not suffer {}from this bias.

This raises the interesting question: since the shear is a non-local
function of the surface density --- so the shear within some region is
determined in part by the surface density outside --- is it possible
to uniquely determine $\kappa$ {}from local measurements?
The answer is affirmative, subject only to the qualification that
one can add an arbitrary constant to $\kappa$. One way to see
this is {}from the expression
for the angular gradients of $\kappa$ in terms of gradients of the shear
(Kaiser 1995):
\begin{equation}	\label{eq:gradkappa}
\vec\nabla\kappa =
\left[
\begin{array}{c}
\kappa_{,1} \\
\kappa_{,2}
\end{array}
\right]
=
\left[
\begin{array}{c}
\gamma_{1,1} + \gamma_{2,2} \\
\gamma_{2,1} - \gamma_{1,2}
\end{array}
\right]
\equiv \bf{u}
\end{equation}
This relation follows directly {}from the expressions
for $\kappa$ and $\gamma_\alpha$ in terms of $\phi$ above, and
`projects out' a certain linear combination of the shear gradients
which are locally determined.
Thus with perfect data we could determine $\kappa$ at any point
relative to the value at some arbitrary reference point simply
by performing a line integral of the observable $\nabla \kappa$.
This also provides an alternative way to visualize the
convolution integral equation (\ref{eq:ks93});
if one averages over radial line integrals {}from points on
some very distant boundary (on which we assume that $\gamma$
and $\kappa$ vanish) to a `target' point $\vec r_0$,
we obtain a two dimensional
integral, and integrating by parts to express everything in terms of
the shear rather than its gradients we obtain equation (\ref{eq:ks93}).
Equation (\ref{eq:gradkappa}) is valid in the weak distortion regime --- the
main focus of this paper --- but can be readily extended into the
strong-distortion regime (Kaiser 1995).

The above argument gives a qualitative insight on how to
account for the bias in the original algorithm and
several groups have exploited the above relations
to develop unbiased surface density estimators (\cite{schneider94};
\cite{kaiser95b}; \cite{schneider95};
\cite{seitzc95};  \cite{seitzs95}).
In this paper, we address the reconstruction problem {}from a variety
of perspectives. First we expand on
some simple alternative solutions to the finite-field problem
where we determine the surface density relative to some well-defined
reference region as sketched in Figures \ref{fig:paths}a,b,c.
In \S\ref{sec:method} we show that in general, if
one uses a line integral averaging scheme then one can formulate an expression
for $\kappa$ relative to the appropriate reference value
as an integral like equation (\ref{eq:ks93}), but with
$\chi_\alpha(\vec r - \vec r_0)
\rightarrow K_\alpha(\vec r;  \vec r_0)$. The benefits of this
approach are twofold: By performing the line integrations and
averaging analytically we are able to construct practical
estimators as simple discrete sums over the background galaxy
shear estimates, so with this method
it is relatively straightforward to obtain rigorous
estimates of the statistical uncertainty in the reconstruction arising
{}from random intrinsic ellipticities. Second, the analysis
reveals how problematic boundary
terms ($\delta$-function terms in the kernel $K_\alpha$)
inevitably arise if one restricts the reference
region geometry to a 1-dimensional line, and how they can be avoided.

Another inversion algorithm has been proposed
(\cite{seitzs95})
that is not only unbiased, but also attempts, in some sense,
to minimize the contribution due to noise. Since the
gradient of the shear is related to $\grad \kappa$, in the absence of
noise, the curl of $\grad \kappa$ should be zero. In practice, noise
adds a rotational part to the surface density gradient and the Seitz \&
Schneider algorithm attempts to minimize this. We show however that
this technique also gives rise to boundary terms which tend to
inflate the noise in the reconstructions and give quite similar results
to other methods that give large weight to galaxies near the boundary.

In \S\ref{sec:fourier} we develop a new discrete Fourier transform based
inverse gradient operator which can be implemented with an fast
Fourier transform (FFT).
Because of periodic boundary conditions, applying the familiar
algebraic inverse gradient operator $1/k$ (or its generalization
for finitely spaced data) does not recover the true surface
density from $\vec \nabla \kappa$.  In fact, for zero padded
data, one recovers precisely the bias inherent in the original
KS93 method.  We show how to remove this bias and derive a Fourier-space
inverse gradient operator which exactly recovers a scalar
field from its gradient.  Unfortunately, this method does not seem
to be optimal in terms of its low frequency noise, but the method
has the advantages that it is simple, fast, and can readily be extended
to apply in the non-linear regime where the data provide one with
a map of $\vec \nabla \log (1 - \kappa)$ (where the low-frequency noise is
not an important issue).

Fourthly, we consider regularized inversion methods for constructing
the surface density. In general terms, this approach arises from fitting a
series of general models for the underlying surface density and attempting
to determine which the most probable model for the true surface density,
given a particular set of observations. This style of approach
leads to two difficulties. First,
if we fit a model with only a few free parameters, then we severely
restrict the reconstructions to some basic forms -- in effect the
reconstructions will reflect
our prejudice for what we expect rather than allowing the data to
determine the surface density.
However if we allow
the model to be very general, then we are essentially
introducing a large number of free parameters and the
inversion become ill-conditioned. We pursue this scheme of many-parameter
inversion in \S\ref{sec:MP} and \S\ref{sec:ML}
and explore two methods to regularize the inversion based on
the maximum likelihood and maximum entropy techniques.

The foregoing analyzes yield several estimators all of which
are unbiased.
We address the question of what estimator is most desirable in practice
in \S\ref{sec:noise}.
They are really precisely equivalent for perfect data, and the
only rationale for choosing one method is how well it performs with
real, noisy data.
For any particular quantity one chooses to measure {}from a
reconstruction the question is well defined and one can objectively
determine the best solution.  The problem is deciding what are
the quantities of interest, which is a somewhat subjective issue.
Here we focus on the low frequency
noise in the reconstructions.  This seems reasonable since the whole
purpose of these modifications to equation (\ref{eq:ks93}) is to cure
the bias which is essentially a low frequency phenomenon.
To explore this we first make simulated reconstructions using
a variety of geometries.  We then make a more objective and
quantitative comparison by seeing how well these methods
perform for particular low-frequency measurements.

In \S\ref{sec:apertures} we derive some useful results
for estimating the mass within an aperture. While this can be
done by performing `aperture densitometry' on the reconstruction,
we show that with a simple modification of formula
for one of the direct reconstruction methods, one can measure the
aperture mass as a
single summation over the shear estimates. This greatly
simplifies the estimation of the statistical uncertainty.
How to do this for a circular aperture has been described
elsewhere (\cite{kaiser95b}) and here we generalize this
to obtain a useful bound on the mass within apertures
of arbitrary shape.

Finally, we consider what appears at first
sight to be a quite different approach to this problem.
Gauss' law in 2-D equates the mass inside a loop and the
integral of the normal gravity around the loop.  Since the
shear is the spatial derivative of the gravity one can
thereby derive an expression for $d \overline \kappa
/ d \ln A$ as a certain integral of the shear around
the boundary.  This provides a generalization of
equation (\ref{eq:tangentialshear}) to non-circular loops. Our hope
was that this would give a different estimator --- the
construction we use in \S\ref{sec:method} being highly
and arbitrarily restricted to averages over straight line
integrals --- but it turns out that the result is exactly
equivalent to our case-b estimator (see below) and so we consign this
to an appendix.

\section{Finite Field Kernels}
\label{sec:method}
\begin{figure}
\myplotfiddle{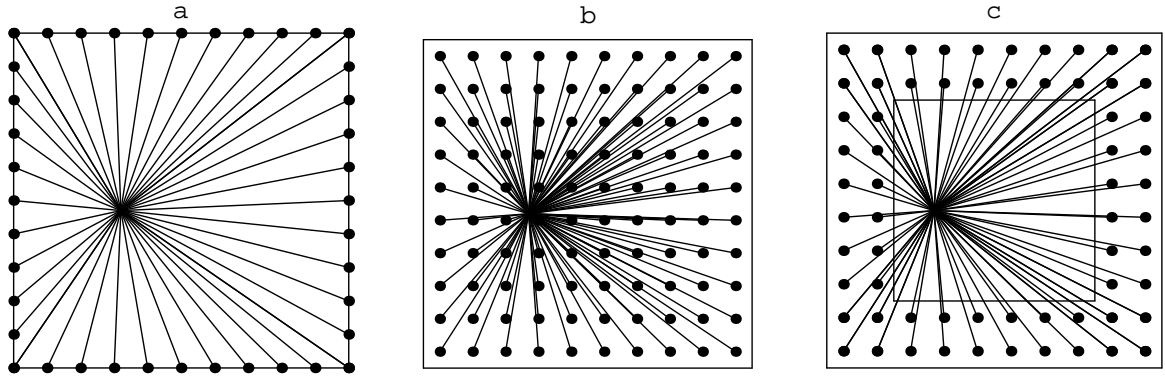}
\caption{Schematic illustration of
line integration paths used to construct the surface
density estimators.  a) Paths for
the estimator of \protect{$\kappa$} relative to
\protect{$\langle \kappa \rangle$} the mean on the
boundary.  b) Paths averaged over if we measure
\protect{$\kappa$} wrt \protect{$\overline \kappa$},
the mean over the whole area surveyed.
c) Hybrid scheme taking reference value \protect{$\tilde \kappa$}
on a strip around the boundary of finite width.}
\label{fig:paths}
\end{figure}

{}From equation (\ref{eq:gradkappa}) we see that it is technically
possible to determine $\kappa$ at any point
relative to the value at some arbitrary reference point simply
by performing a line integral of the observable $\nabla \kappa$.
Clearly then, one could recover $\kappa$ by simply averaging
$\int dl \cdot \nabla \kappa$ over radial lines {}from the boundary
of the observed region, as suggested by
Kaiser {\it et al.\/} (1995) and illustrated schematically in Figure
\ref{fig:paths}a, and an algorithm to implement
such a scheme was developed by Schneider (1994).

Schneider's algorithm evaluates the line integrals
numerically. To do this the shear estimates are first binned into
grid of cells and then smoothed.  The gradients of the shear
(and hence gradients of $\kappa$) are approximated by discrete
differences on the smoothed shear field and $\nabla \kappa$ is
then numerically integrated along a family of lines extending {}from
each point on the plane to the boundary. Now the geometry for the line
integrals is actually slightly different {}from that shown in
Figure \ref{fig:paths}a, but shares the key property that
the boundary points are the same for each target
point so, for perfect data at least, this procedure should generate
an exact and unbiased reconstruction.

The bias in the low spatial frequency behavior in the KS93 estimator
is a real problem --- especially if one is interested
in the mass profile in the outer parts of cluster --- and
Schneider's algorithm is an interesting attempt
to avoid this, but it has one somewhat disturbing feature.
The line integrals of gradients of the shear
give rise to integrals of the shear plus boundary terms.
In the averaging over lines, the former become two dimensional
integrals which are very similar in form to the convolution
integral in equation  (\ref{eq:ks93}), and are relatively
straightforwardly implemented
as discrete sums over galaxies. The
latter, however, become line integrals of the shear around the
perimeter. As we will see, these make an important contribution,
especially to the low frequency components of the reconstruction,
but they are very difficult to estimate reliably.
The problem is that if the shear is smoothed then its value
at the boundary will tend to be biased.
On the other hand, if the smoothing radius is taken
to be very small then the boundary shear must be estimated
by averaging over a small number of galaxies and this will
inflate the statistical uncertainty in the reconstruction.

As discussed elsewhere (\cite{kaiser95b}) there is a simple yet general
relation between the mean tangential shear around a circular loop
and the mean surface density within that loop.
Defining the tangential shear as
$\gamma_T = - (\gamma_1 \cos 2 \varphi + \gamma_2 \sin 2 \varphi)$
--- it measures the stretching of the galaxies along the
loop --- and the mean interior surface density as
$\overline \kappa = M / A$, one finds
\begin{equation}	\label{eq:tangentialshear}
\langle \gamma_T \rangle =
-{d \overline \kappa \over d \ln A}
\end{equation}
This is easily derived {}from Gauss' law
(see appendix \ref{sec:aperturesappendix}), and
can been used to estimate the mass within a circular aperture
(\cite{fahlman94}). If we simply evaluate
$\int d\ln A \langle \gamma_T \rangle$ out to
some radius $r$ we obtain the surface density at the origin
minus $\overline \kappa(r)$, the mean over the disk.
Now by definition $\overline\kappa = (2/r^2)\int dr r
\langle \kappa \rangle$, where $\langle \kappa \rangle$ denotes
the mean of $\kappa$ {\sl on\/} the loop, {}from which it
follows that $\overline \kappa = \langle \kappa \rangle -
d \overline \kappa / d \ln A$.  Comparing with equation
(\ref{eq:tangentialshear}) we can see that if we were to change
our reference surface density {}from $\overline \kappa$
to $\langle \kappa \rangle$ the only effect is to introduce
the pure boundary term $\langle \gamma_T \rangle$.

Our analysis shows that it
is technically impossible to measure the surface density relative to
the mean on a boundary of infinitesimal width --- which is perhaps
hardly surprising --- but allows the possibility that one measure
$\kappa$ relative to a well defined reference strip of thin but finite
width, which would give an approximation to Schneider's method
but without the bias, or any similar alternative such as
Figure \ref{fig:paths}b.

In this section, we
construct expressions for $\kappa$ as averages of line integrals
of $\nabla \kappa$ {}from a set of points uniformly distributed
over some reference region
for the various geometries illustrated schematically in Figure
\ref{fig:paths} (although the square geometry is only illustrative and the
analysis below applies general survey geometries).
Then, using equation (\ref{eq:gradkappa}), we obtain explicit expressions for
$\kappa$ as some integral over the shear field which
can then readily be converted in \S\ref{sec:estimators} to form practical
estimators of $\kappa$ given a set of shear estimates.

\subsection{case a:}

We first construct an estimator for $\kappa - \langle \kappa \rangle$
where $\langle \kappa \rangle$ is the mean surface density on the
boundary of the data region.
Place the origin of coordinates at the `target point' $\vec r_0$
where we wish to
measure $\kappa$.  Let the boundary of the region on which we
have data be $\vec p(\varphi)$, parameterised by azimuthal
angular coordinate $\varphi$. We will assume that the entire boundary
is visible {}from any interior point where we will attempt to estimate
$\kappa$.
The mean of $\kappa$ on the
boundary is $\langle \kappa \rangle = \int d \varphi W(\varphi) \kappa$
where $W \equiv |\vec t| / L$, with
$L \equiv \int d \varphi |\vec t|$ the length of the perimeter, with
tangent vector
$\vec t \equiv d \vec p / d \varphi$.  Taking the average over
radial paths to the boundary as in Figure \ref{fig:paths}a we find
\begin{equation}	\label{eq:kappahat0}
\kappa - \langle \kappa \rangle
= -\int d \varphi W(\varphi) \int\limits_0^{p(\varphi)} dr
\vec r \cdot \vec \nabla \kappa / r
= -\int d^2 r W \vec r \cdot \vec \nabla \kappa / r^2
\end{equation}
Substituting equation (\ref{eq:gradkappa}) for $\nabla \kappa$ and
integrating by parts to express everything in terms of $\gamma$
rather than its derivatives we obtain
\begin{equation}	\label{eq:kappahat1}
\kappa - \langle \kappa \rangle
= \int d^2 r K_\alpha \gamma_\alpha -
{1\over L} \oint dl Y_\alpha \gamma_\alpha
\end{equation}
where
\begin{equation}	\label{eq:K1}
K_\alpha = 2 {\cal G}_\alpha (W / r^2)
\end{equation}
where we have defined the differential operator
\begin{equation}	\label{eq:Gdefinition}
{\cal G}_\alpha \equiv {1\over 2}
\left[
\begin{array}{c}
x {\partial \over \partial x} -
y {\partial \over \partial y} \\
y {\partial \over \partial x} +
x {\partial \over \partial y}
\end{array}
\right]
\end{equation}
and where the loop integral is taken around the perimeter
with
\begin{equation}	\label{eq:Y1}
Y_\alpha =
\left[
\begin{array}{c}
(p_x t_y + p_y t_x) / p^2 \\
(p_y t_y - p_x t_x) / p^2
\end{array}
\right]
\end{equation}
With the origin placed at the center of a circular survey
region we recover the KS93 kernel and, as expected,  the surface
term is just the mean tangential shear on the boundary.  Equations
\ref{eq:kappahat1}, \ref{eq:K1}, \ref{eq:Y1} provide the
generalization for points off the axis and/or non-circular
survey geometries.

This is clearly very similar to Schneider's method.
Let us clarify this.  His method consists of two
sequential operations on the data.  The first is the
smoothing of the shear field to produce a fine grid
of $\gamma_\alpha$ values.  The second consists of differencing
these binned values, grouping the various components together
to make $\nabla \kappa$ and then integrating. Let us ignore the
smoothing for now.
Now the second step is just some linear operation on the input grid of shear
values, and could be written out as a linear sum: $\kappa =
\sum c_\alpha \gamma_\alpha$.  What we have done above is to explicitly
calculate the coefficients $c_\alpha$ in this sum. Now in fact Schneider
measures $\kappa$ relative to a slightly different average on the
boundary, and uses a slightly different geometry
for the rays along which the line integrals are performed,
but these are irrelevant details.
The worrying thing which emerges {}from our analysis is the presence
of the boundary term, which means that the grid points on the
boundary receive a very high weight in this sum.  Now as we
have seen, the contribution of this term to $\hat \kappa$
is on the order of the mean shear on the boundary, which is
roughly equal to $\overline \kappa$, so clearly the contribution
to the low spatial frequency components in the reconstruction
{}from this surface term is substantial, and since the goal here was to fix
the bias in
the low-frequency behavior of KS93, it is important that
the method get this right.  It is hard to see how this can be, since if the
shear field is smoothed then the value on the boundary will be
biased.  One the other hand with a small smoothing length the
boundary term must be calculated by averaging over a small number of galaxies
and this will inflate the statistical uncertainty.

\subsection{case b:}

\begin{figure}
\myplotone{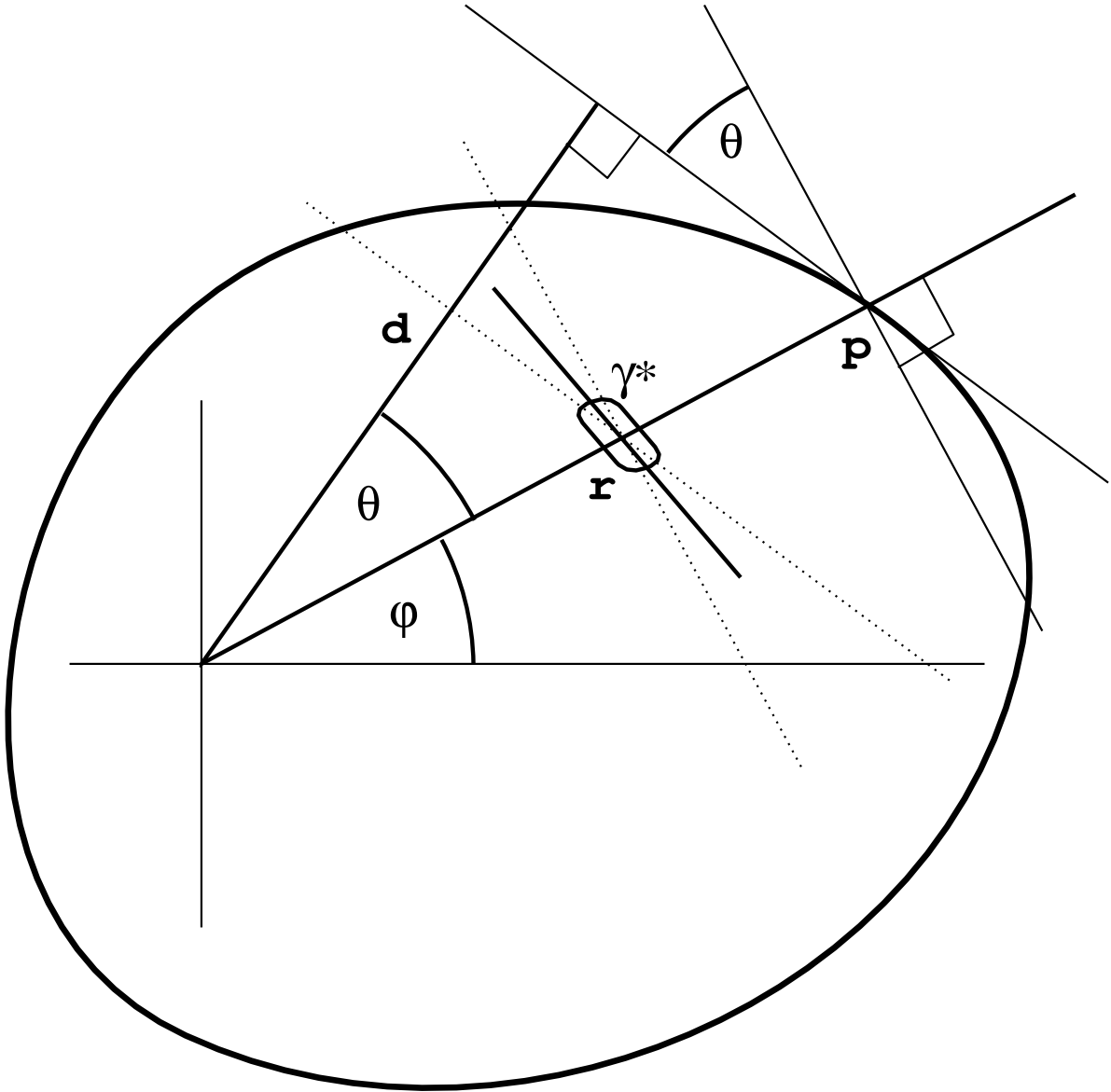}
\caption{The heavy loop represents the boundary of the
region on which we have data.  The target
point lies at the origin of coordinates; \protect{$\vec p$} is the point
on the boundary lying behind the measurement point \protect{$\vec r$}.
The dotted lines are parallel to the normal to \protect{$\vec p$} and
the tangent vector at $\vec p$ respectively, and the ellipse
shows the orientation of the unit polar \protect{$\gamma^*_\alpha$} whose
major axis is rotated anticlockwise {}from the vertical by an
angle \protect{$\varphi + \theta / 2$}.  Also shown is \protect{$d$}, the
perpendicular distance {}from the origin to the tangent vector.}
\label{fig:angles}
\end{figure}

Let us now construct instead an estimator of $\kappa - \overline \kappa$;
where the zero-point is now set to be
$\overline \kappa = \int d^2r \kappa / A$,
with $A = \int d^2 r$.  Taking the average over radial
lines to the origin {}from each point within the survey region, as
illustrated in Figure \ref{fig:paths}b, we have
\begin{equation}	\label{eq:kappahat2}
\kappa - \overline \kappa =
{- 1\over A} \int d \varphi \int\limits_0^{p(\varphi)} r dr
\int\limits_0^r dr' \vec r' \cdot \vec \nabla \kappa(\vec r') / r'
=  {- 1\over 2A} \int d \varphi \int\limits_0^{p(\varphi)} r dr
(p^2/r^2 - 1) \vec r \cdot \vec \nabla \kappa
\end{equation}
This is very similar in form to equation \ref{eq:kappahat0}, but now
the function $W$ is replaced by the function
$(p^2 - r^2) / 2 A$
which vanishes on the
boundary, so substituting {}from equation (\ref{eq:gradkappa}) and
integrating by parts we obtain
\begin{equation}	\label{eq:kappahat3}
\kappa - \overline \kappa
= {1\over A} \int d^2 r K_\alpha \gamma_\alpha
\end{equation}
where now
\begin{equation}	\label{eq:K2}
K_\alpha = {\cal G}_\alpha (p^2 / r^2)
\end{equation}
but no boundary term.

A geometric picture of this kernel may be obtained as
follows: In polar coordinates ($r,\varphi$) we find
${\cal G}_\alpha = (1/2) R_{\alpha\beta}(2 \varphi)
\{ r \partial / \partial r,
\partial / \partial \varphi \}$ where
$R_{\alpha\beta} = $$\{\{\cos,$$-\sin\}$,$\{\sin,$$\cos\}\}$
is the 2-dimensional rotation matrix and, since
$p = p(\varphi)$, we have
$K_\alpha = - (p / r^2) R_{\alpha \beta}(2\varphi)
\{p, p'\}$ with $p' \equiv d |p| / d \varphi$.  Now {}from
inspection of
Figure \ref{fig:angles} one can see that $p' = p \tan \theta$,
and the perpendicular
distance {}from the origin to the tangent is $d = p \cos \theta$,
so we find  $K_\alpha = p^3 \gamma^*_\alpha / r^2 d$
where the `unit polar' is defined to be
\begin{equation}
\gamma_\alpha^* \equiv
-\left[
\begin{array}{c}
\cos 2 (\varphi + \theta / 2)
\\
\sin 2 (\varphi + \theta / 2)
\end{array}
\right]
\end{equation}
and hence
\begin{equation}	\label{eq:kappahat4}
\kappa - \overline \kappa
= {1\over A} \int d^2 r  {p^3 \over  r^2 d} \gamma^*_\alpha \gamma_\alpha
\end{equation}
This is very similar to equation (\ref{eq:ks93}).
Both  equations `project out' a
particular component of the shear.  In  equation (\ref{eq:ks93}) this is
simply the tangential shear; i.e.~the stretching perpendicular
to the line {}from the origin to the measurement point. In equation
(\ref{eq:kappahat4}) the component projected out measures the
stretching along the direction which bisects the normal to
the vector $\vec p$ and the boundary tangent direction.
The orientation of $\gamma^*$ is shown in Figure \ref{fig:angles}.

\subsection{case c:}

We can now readily
generalize this analysis to the case where the reference
region does not fill the entire data region or has some complicated geometry
so that lines {}from the target point may pass in and out of the
reference region, perhaps many times.
If we think of the ends of the
lines in Figure \ref{fig:paths}b,c as lying on a unit spaced
regular 2-dimensional grid (and let the unit length be very small) then
$A(\kappa - \overline \kappa)$ is just the
sum over line integrals. Clearly, if we excise part of the reference region
say, we must simply remove the contribution {}from
all the paths which originate in the excised region.  In the general
situation a line {}from the target point (with some given azimuthal angle
$\varphi$) will pass through the boundary of the reference region (which need
not be simply connected) at points $p_1 \ldots p_n$.  Let the points
be sorted in order of increasing distance so the most distant point is $p_n$.
The mathematical expression of the subtraction described above is
the generalized kernel
\begin{equation}	\label{eq:generalK}
K_\alpha = \sum_i S_i \Theta(r - p_i) {\cal G}_\alpha (p_i^2/ r^2)
\end{equation}
where $S_n = 1$, and $S_{n-1} = - S_n$ and where $\Theta(x)$ is the
Heaviside function.  The surface density is then given by equation
(\ref{eq:kappahat3})
as before, but with $A$ now the area of the reference region.

\begin{figure}
\myplotoneb{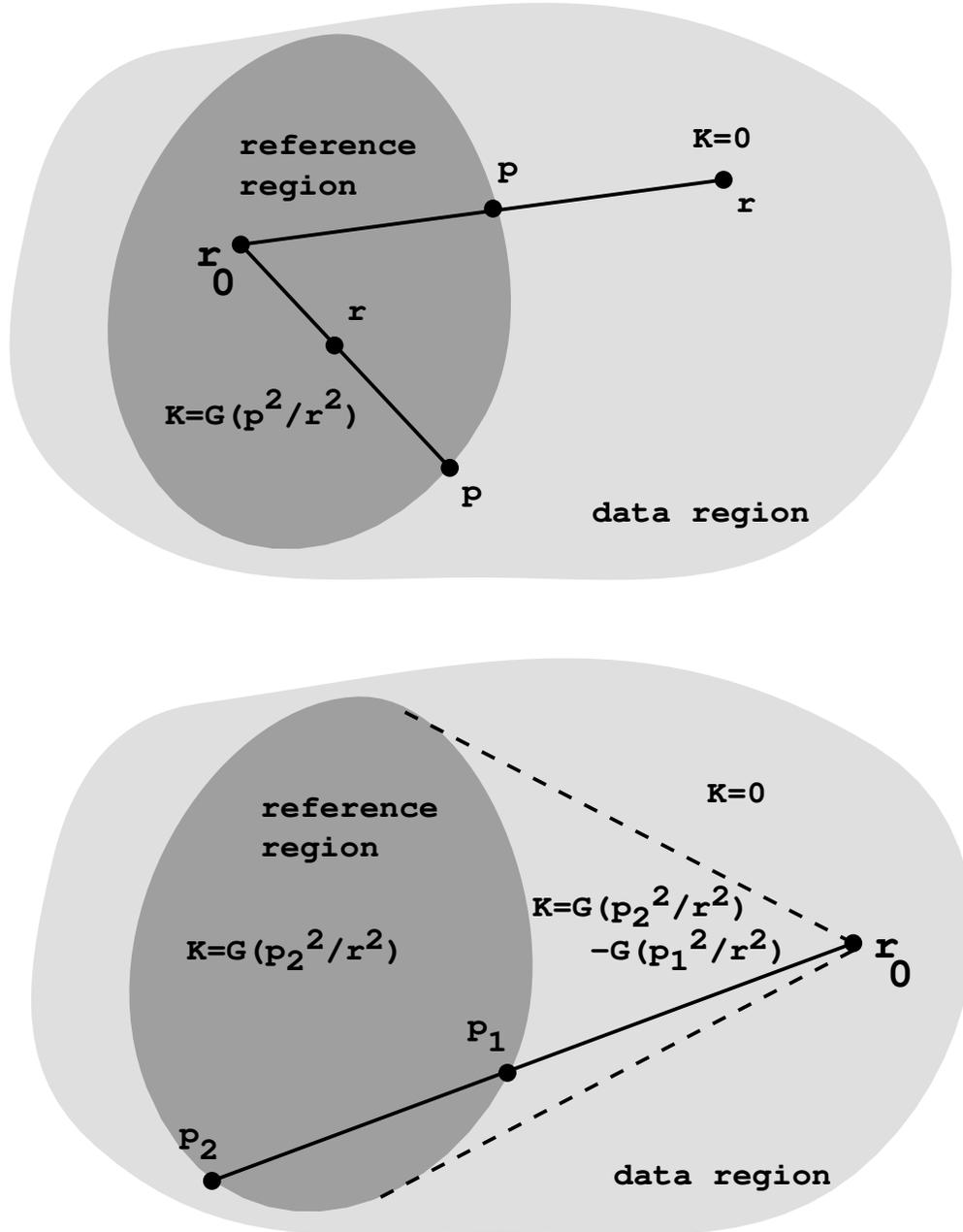}
\caption{Generalization of case b when the reference region
only partially fills the region on which we have data.
The form of the kernel is shown for the
various possible cases when the point in question
lies inside the reference region (upper panel) or outside
it (lower panel).}
\label{fig:general}
\end{figure}

This is illustrated in Figure \ref{fig:general} for the case
of a simply connected reference region which only partially
fills the region on which we have data.
If the target point $\vec r_0$ lies inside the
reference region (upper plot) then the line through $\vec r$
crosses the boundary just once and we have
\begin{equation}
K_\alpha =
\left\{
\begin{array}{c}
{\cal G}_\alpha (p^2 / r^2)\\
0
\end{array}
\right.
{\rm for}
\left\{
\begin{array}{c}
r < p\\
p < r
\end{array}
\right.
\end{equation}
If the target point lies outside the reference region then either
the line {}from $\vec r_0$ though $\vec r$ misses the reference
region completely (in which case $K_\alpha = 0$) or it
crosses it twice at $\vec p_1$ and $\vec p_2$ as illustrated in the
lower panel, and we then have.
\begin{equation}
K_\alpha =
\left\{
\begin{array}{c}
{\cal G}_\alpha (p_2^2/ r^2) - {\cal G}_\alpha (p_1^2/ r^2) \\
{\cal G}_\alpha (p_2^2/ r^2)\\
0
\end{array}
\right.
{\rm for}
\left\{
\begin{array}{c}
r < p_1\\
p_1 < r < p_2\\
p_2 < r
\end{array}
\right.
\end{equation}

We can now easily construct the kernel for case-c, where
we have two nested boundaries; we simply subtract the
kernel calculated as above for the inner boundary {}from that
for the outer.  This also allows the possibility that
the outer boundary might lie inside the actual data
boundary. Case-c approaches case-a in the limit that the reference strip is
infinitesimally thin. If we use a narrow but finite width
strip we should obtain something very similar to case-a, but
without the bias inherent in Schneider's method.

These formula were derived for the special case where
the spatial origin coincides with target point.
For a general choice of origin we have
\begin{equation}
\kappa(\vec r_0) - \overline \kappa
= {1\over A} \int d^2 r K_\alpha(\vec r; \vec r_0)
\gamma_\alpha(\vec r)
\end{equation}
where, for case-b for example, $K_\alpha(\vec r; \vec r_0) =
{\cal G}_\alpha(p^2(\vec r,\vec r_0) / (\vec r - \vec r_0)^2)$,
where we have explicitly shown the dependence of the boundary
point on the target and measurement points.

\subsection{Example Kernels}

\begin{figure}
\myplotoneb{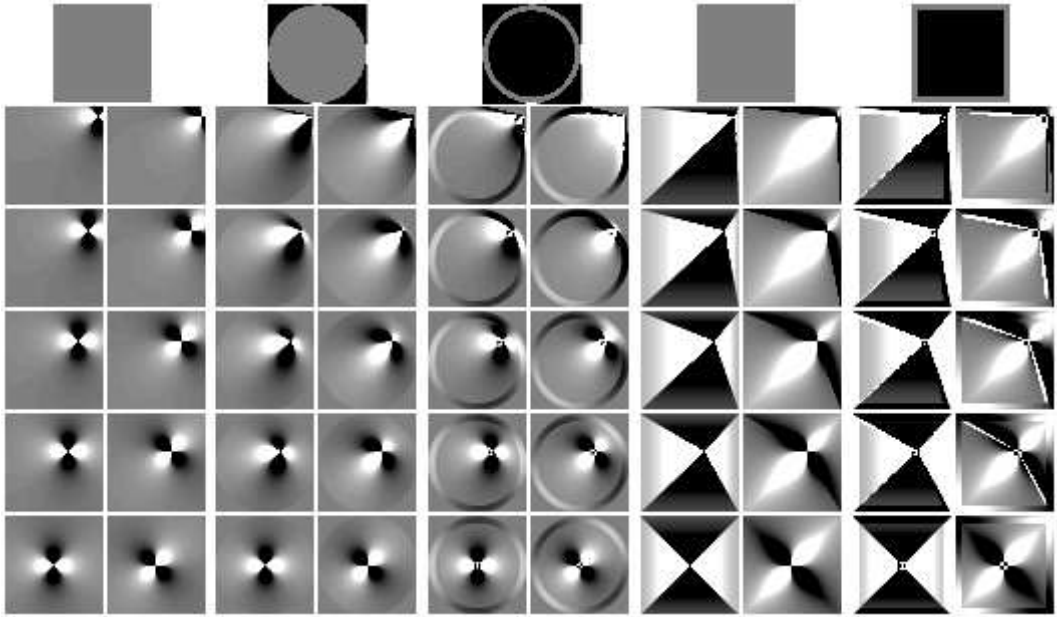}
\caption{Plots showing the kernels for various reference
region geometries and for various target positions.  The
leftmost pair of columns give the KS93 kernels \protect{$\chi_1$} and
\protect{$\chi_2$}.  The next pair show the case-b kernel for a
circular survey geometry and the middle pair show the case-c kernel.
The two pairs of columns on the right show
the case-b and case-c kernels for rectangular geometry.
The case-c kernels are very much like their
case-b counterparts, save for the enhanced value within the
reference annulus itself.  As the reference strip becomes thinner
this becomes stronger.}
\label{fig:kernels}
\end{figure}

An illustrative and interesting case is that of the case-b
kernel for a rectangular survey geometry.
Erecting lines {}from the target point (the point at which we hope to
measure $\kappa$) to the corners of
the rectangle divides it into N,S,E and W quadrants.
{}From equation (\ref{eq:kappahat4}) we find,
in the N and S sectors respectively,
$K_1 = (Y \mp y_0)^2 / (y - y_0)^2$ where $Y$ is
half height of the rectangle, and
$K_2 = -(x - x_0) K_1 /(y - y_0) $.
Similarly, for galaxies in the E and W
sectors we have $K_1 = - (X\mp x)^2 / (x - x_0)^2$ and
$K_2 = (y - y_0) K_1 / (x - x_0)$ where $X$ is the half-width.
This kernel, is shown in the Figure \ref{fig:kernels}
for various target positions.
The pattern
is qualitatively similar to the KS93 kernel, which is also
shown for comparison, but now depends in a non trivial way on
$\vec r_0$.  Note also the discontinuity along the divisions
between the N,S,E,W sectors.  As we will see, these give rise
to spurious, but essentially harmless,
linear features in the reconstruction.

Also shown in Figure \ref{fig:kernels} are the kernels for the case of
circular reference region lying within the survey region (assumed square for
simplicity).
If we define $\mu = \vec r \cdot \vec r_0 / r r_0$,
and $Z = \sqrt{R^2 / r_0^2 + \mu^2 - 1}$ where $R$ is
the radius of the survey and let $Q = Z - \mu$ then
for case-b, setting $r' = r - r_0$ and dropping primes
we have
\begin{equation}
K_\alpha =
- {r_0^2 Q^3 \over Z r^4}
\left[
\begin{array}{c}
x^2 - y^2\\
2 x y
\end{array}
\right]
+ {r_0 Q^2 \over Z r^3}
\left[
\begin{array}{c}
x x_0 - y y_0\\
x y_0 + y x_0
\end{array}
\right]
\end{equation}
These formulae are only valid if both target and observation point
lie within the survey disk, and must be modified if the target
or observation point lie outside as discussed above.

The case-b and case-c
circular kernels are in each case very similar, and the main difference
is just the extra term around the annulus.  As one makes the
width of the reference region smaller the kernel in the annulus increases
in inverse proportion and so, in the limit of a very thin annulus we
end up giving infinite weight to the infinitesimal number of
galaxies lying in the annulus.

\subsection{Practical Estimators}
\label{sec:estimators}

Given a catalogue of background galaxy ellipticities
$e_\alpha$ which, suitably
defined (e.g.~\cite{ksb95}),
give a set of estimates of $\gamma_\alpha$ at their
locations $\{\vec r_g\}$, we can form the estimator
\begin{equation}
{\hat \kappa}(\vec r_0)
= {A_{\rm tot} \over A}
{\sum_g w_g K_\alpha(\vec r_g; \vec r_0) e_\alpha \over \sum_g w_g}
\end{equation}
whose expectation value is just $\kappa(\vec r_0)$ minus
the appropriate reference value, and where $\{w_g\}$ are a set
of weights that we can tune to optimize performance.
Alternatively, one could bin the galaxies on some grid, calculate
the mean shear estimate ${\hat \gamma}_\alpha$
for each bin, and then evaluate
$\int d^2r\; K_\alpha {\hat \gamma}_\alpha$
as a discrete sum with appropriate weights. These two approaches are
equivalent in the limit of small bin size.
For uniformly sampled points (say on a fine grid)
the weights should clearly all be equal.  Given some
random realization of background galaxy positions the constant
weight estimator is no longer ideal: the kernel we
are effectively applying is $n K_\alpha / \overline n$ which
differs {}from the true kernel because of the fluctuating density
of the background galaxies. A uniform weighting
scheme therefore introduces a {\sl multiplicative\/} element
of noise, which can be removed by
weighting each galaxy in inverse proportion to the local
density of neighbors.
However, the {\sl additive\/} error arising {}from the random
intrinsic ellipticities will consequently
increase.  The optimum weight therefore depends on signal
to noise, with constant weight preferred in low signal
to noise situations.  For typical clusters the distinction between
these is minor as to get into a regime of high S/N one really
has to look at large-scale features and then the Poissonian
fluctuations in the background galaxy number density
are small. Henceforth we will consider the uniform weight estimator
\begin{equation}	\label{eq:kappahat5}
{\hat \kappa}(\vec r_0)
= {1 \over \overline n A}
\sum_g K_\alpha(\vec r_g; \vec r_0) e_\alpha
\end{equation}

What about allowance for contamination by cluster galaxies?
If one assumes that they simply add a randomly oriented component
to the background galaxy population then one should
simply use equation (\ref{eq:kappahat5}) as it stands but with $\overline n$
determined away {}from the cluster.  If, however, the
light {}from the cluster galaxies masks a
significant fraction of the background galaxies one should
boost the weight given to the surviving galaxies by an
appropriate amount.

Note that all the the kernels are
strongly divergent at $r \rightarrow 0$, naively
suggesting a logarithmic divergence in the kernel integral if we simply
count powers of $r$. However, this divergence is not real.
Consider the case b kernel given in equation (\ref{eq:K2}).
If we make a Taylor expansion of $\gamma_\alpha$ about the origin
then it is only the zeroth order term which threatens to produce
a divergence, gradients or higher order terms being clearly convergent.
For $\gamma_\alpha$ = constant, we can integrate equation (\ref{eq:kappahat3})
by parts.
If we restrict the integration to some patch around the origin
we find
\begin{equation}
\begin{array}{c}
\int d^2 r K_\alpha \gamma_\alpha =
\int dy(x \gamma_1 + y \gamma_2) \left[p^2/r^2\right]_{x_1}^{x_2}\\
+ \int dx(x \gamma_2 - y \gamma_1) \left[p^2/r^2\right]_{y_1}^{y_2}
\end{array}
\end{equation}
This is a boundary term which vanishes identically if we take the
boundary to be $p^2 / r^2$ = constant. Thus, if we define the
integral in equation (\ref{eq:kappahat3})
as the limit as $\epsilon \rightarrow 0$
of $\int_{r>\epsilon p}d^2r \ldots$ it is then clearly non-divergent.

A closely related issue is the question of noise in
the surface density estimator given in equation
(\ref{eq:kappahat5}) arising {}from the divergence of the kernel.
The reconstruction is the sum of random `shots' convolved with the
kernel, now considered a function of $r_0$.
Clearly,  if we make an
unsmoothed reconstruction we will find divergent `butterfly' patterns
at the location of each galaxy.  In the case of the KS93
kernel it is easy to see that this noise is a purely high
frequency phenomenon and goes away on smoothing the final density
field.  The reason for this
is that the kernels $\chi_1$, $\chi_2$ vary as $\cos 2 \varphi$,
$\sin 2 \varphi$ so the coupling to long wavelength Fourier modes is
highly suppressed.  Here essentially the same is true, but we have to be
a little careful. In the KS93 estimator one can simply soften the
divergence by replacing $1/r^2 \rightarrow 1 / (r^2 + r_s^2)$
for some tiny softening length which merely suppresses overshoot
in the computer. Were we to do that here, there would be a divergent
coupling to low frequency modes because of the asymmetry
of the kernel pattern which, in reality, would be cut-off
by the finite pixel size in the reconstruction.  The answer is
to soften the kernels by multiplying by some function
$W(r/\epsilon p)$ where $\epsilon$ is a small parameter and
where $W(x) \rightarrow 0$  for $x \ll 1$ and
$W(x) \rightarrow 1$  for $x \gg 1$ (the function used in the
simulations shown below was $1 - \exp(-x^3)$).
In principle, the long wavelength behavior of the reconstruction
should be independent of the parameter $\epsilon$ provided
it is `sufficiently small'.  A slight technical problem with this is
that the scale length of the softening becomes small for those galaxies
near the boundary and so quite a small pixel size is required in
order to resolve this properly.

\subsection{Minimizing Noise Contributions}
\label{sec:SS}
Another way of approaching the problem is to notice that {}from
equation (\ref{eq:gradkappa}), $\grad \times \grad \kappa$ = 0 in the
absence of noise. Hence for real, noisy data, a sensible suggestion
is that the rotational component of the gradients of the shear field should,
in some sense, be minimized. Seitz and Schneider (1995) show that by adopting
two reasonable assumptions:
1) the rotational component should vanish if the shear is a gradient field and
2) there is no systematic rotation over the data field, then there is
a unique specification for decomposing the gradient of $\kappa$ into
a gradient and rotational part. The details are involved but the
main result is that this gives an expression for
$\kappa$ in the familiar format as a convolution of the gradients of
the shear with a kernel:
\begin{equation}
\kappa(\vec{r_0} ) - \bar{\kappa} = \int d^2r
			K_\alpha(\vec{r};\vec{r_0}) u_\alpha( \vec{r} )
	\label{eq:ssconvolution}
\end{equation}
where $u_\alpha( \vec{r} )$ is defined in equation (\ref{eq:gradkappa}).

The problem arises when we attempt to express this with respect to the
shear, instead of its gradients. In solving for the kernel, Seitz and
Schneider are effectively solving a Neumann problem, where the kernel
is a gradient vector field with
$K_\alpha(\vec{r};\vec{r_0}) n_\alpha(\vec{r}) = 0$ on the boundary, and where
${\bf n}(\vec{r})$ is the unit normal to the boundary.
Integrating equation (\ref{eq:ssconvolution}) by parts, we obtain an
estimate for
the surface density as a convolution of a kernel with the shear estimates
{\em plus} a boundary term, much like our case-a estimator.

One might expect then, that the limitation of this method is
very similar to the case-a estimator
(or thin reference region case-c): to estimate
the gradients of the shear field, it is necessary to introduce some smoothing.
This will bias the estimate near the boundary, or inflate the noise
if the smoothing is reduced. The latter is particularly problematic.
The presence of a boundary term tends to give very large weight to a
few galaxies near the boundary and this introduces a worrisome low frequency
fluctuation in the reconstructions - yet it was just this feature we are
attempting to reduce in considering alternatives to the original
KS93 estimator. We return to this problem more quantitatively
in \S \ref{sec:noisepower}.

\section{Direct Fourier Reconstruction}
\label{sec:fourier}

We can construct a binned
estimate of $\vec \nabla \gamma_\alpha$ by binning $\gamma_\alpha$
values on regular grid and then discrete differencing. Hence we can
obtain binned estimates for $\vec \nabla \kappa$,
via equation (\ref{eq:gradkappa}).
The finite field methods have attempted to determine $\kappa$ by averaging
over line integrals. An alternative method to determine $\kappa$
is obtained by using Fourier transforms. The clear
advantage here is that the gradient operator in real-space becomes
algebraic in k-space. An added bonus is that fast Fourier transforms
are computationally cheap to perform so the inversion is
easy to do numerically. We display such an implementation scheme here.

Consider first, for clarity, a 1-dimensional process, and
imagine we have some discrete process $f[n]$ (this might
be a discretely sampled continuous field: $f[n] = f(n \Delta x)$)
and make the discrete difference
\begin{equation}
\eta[n] = f[n+1] - f[n]
\end{equation}
Now we can construct the Fourier transforms (FT)
of $f$, $\eta$, for any samples
of finite length $N$:
\begin{equation}
\begin{matrix}{
\tilde f[k] = \sum\limits_{n=1}^{N} f[n] e^{2 \pi i n k / N} \cr
\tilde \eta[k] = \sum\limits_{n=1}^{N} \eta[n] e^{2 \pi i n k / N}
}\end{matrix}
\end{equation}
The question is, how are $\tilde f[k]$ and $\tilde \eta[k]$ related?
In particular, if we are given the latter, can we recover the former?

Now for continuous fields $f(x)$, $\eta(x) = d f / dx$, the derivative operator
in Fourier space is just multiplication by $ik$, so we recover the FT of $f$
by dividing FT of $\eta$ by $ik$.  This is not defined for $k = 0$, so this
tells us
we can not recover the DC component -- again, this is not problematic as the
DC level can not be determined from shear measurements alone in any case.

It is well known that with for discrete FTs the discrete difference operator
is multiplication by $e^{-2\pi i k / N} - 1$ rather than $ik$, but there is a
little more to it than that because of the periodic boundary conditions:
if one simply multiplies $\tilde f[k]$ by $e^{-2\pi i k / N} - 1$ then one
obtains the FT of the derivative of $f[k]$ but taken with periodic boundary
conditions which differs {}from the real discrete derivative of $f[k]$ at
the end points.

The actual relation between $\tilde f[k]$ and $\tilde \eta[k]$ is easily
obtained:
\begin{eqnarray}
\tilde \eta[k] &= &\sum\limits_{n=1}^{N} \eta[n] e^{2 \pi i n k / N} =
\sum\limits_{n=1}^{N} (f[n+1] - f[n]) e^{2 \pi i n k / N} \nonumber \cr
&=& \sum\limits_{n=2}^{N+1} f[n] e^{2 \pi i (n - 1) k / N} -
\sum\limits_{n=1}^{N} f[n] e^{2 \pi i n k / N} \nonumber \cr
&=& f[N+1] - f[1] + (e^{-2\pi i k / N} - 1)
\sum\limits_{n=1}^{N} f[n] e^{2 \pi i n k / N} \nonumber \cr
&= &\sum\limits_{n=1}^{N} \eta[n] + (e^{-2\pi i k / N} - 1) \tilde f_k
\nonumber \cr
&= &\tilde \eta[0] + (e^{-2\pi i k / N} - 1)  \tilde f[k] \nonumber
\end{eqnarray}
so the solution of our inversion problem is
\begin{equation}
\tilde f[k] = {\tilde \eta[k] - \tilde \eta[0] \over e^{-2\pi i k / N} - 1}
\end{equation}

As before, this is not defined for $k = 0$, but also
differs {}from usual formula $\tilde \eta[k] / (e^{-2\pi i k / N} - 1)$
for $k \ne 0$.
One way to implement this is simply to modify $\eta[N]$,
the last element of the
$\eta[n]$ array, to be minus the sum of the other elements.  That is, we
construct
\begin{equation}
\eta'[n] = \eta[n] - \delta_{nN} \sum\limits_{n' = 1}^N \eta[n'] =
\eta[n] - \delta_{nN} \tilde \eta[0]
\end{equation}
which then has discrete
transform $\tilde \eta'[k] =  \tilde \eta[k] - \tilde \eta[0]$.

The above analysis can be extended to the case of 2
dimensions, as is applicable for our problem. We can estimate directly
{}from the data the two partial derivatives
$\kappa_{,x}$ and $\kappa_{,y}$ and, in principle, we can use either
to estimate $\kappa$. Fourier transforming, we
obtain two measures of $\kappa$:
\begin{eqnarray}
\tilde \kappa_a[k_x,k_y] & = & \frac{\widetilde{\kappa_{,x}}[k_x,k_y]}
			{ e^{2\pi i k_x/N_x} - 1 } \nonumber \\
\tilde \kappa_b[k_x,k_y] & = & \frac{\widetilde{\kappa_{,y}}[k_x,k_y]}
			{ e^{2\pi i k_y/N_y} - 1 }
\end{eqnarray}

As with the KS93 method expressed in Fourier space these two
estimators have noise which varies strongly with direction
in $k$-space, so
we combine these with weights $w_a = k_x^2 / k^2$, $w_b = k_x^2 / k^2$,
which, for high frequencies at least, gives the optimal combination:
\begin{equation}
\tilde\kappa[k_x,k_y] = \frac{k_x^2}{k^2} \tilde\kappa_a[k_x,k_y] +
	\frac{k_y^2}{k^2} \tilde\kappa_b[k_x,k_y]
\end{equation}

It is very simple to implement  this technique.  One simply bins the
shear estimates onto a grid (this can be very fine as the FFT is
very fast, and the fact that most of the grid cells are actually empty
causes no problem).  We then apply the two discrete difference
operators to this grid of shear values and then combine the
gradients according to equation (\ref{eq:gradkappa}) which, for a very fine
grid, produces an image (or rather pair of images; one for each component
of the gradient) which is largely zero and has a little L-shaped pattern
at the location of each galaxy and we then simply apply the FFT,
calculate $\tilde \kappa$ according to the two equations directly above,
and then inverse transform.  The results from this are very hard
to distinguish visually from e.g. our other best methods, but, as
we shall see, the noise power analysis suggests that the low-frequency
performance of this method is somewhat worse than the best methods.

Doing the inversion in Fourier space also leads to the attractive
possibility to perform the non-linear lensing inversion. It is possible
to construct the quantity $\vec{\nabla}\log(1-\kappa)$ from
observable quantities (Kaiser 1995), in exact analogy with equation
(\ref{eq:gradkappa}). Thus simply replacing $\kappa$ by
$\log(1-\kappa)$ in the equations above gives a prescription for
doing the inversion, even in the non-linear regime.

\section{Regularized Inversion Methods}

{}From the finite field `direct reconstruction' methods,
we obtain the $\kappa$ estimate as a simple convolution of the data with
some geometric kernels.
This is guaranteed to be numerically stable and unbiased and
has well defined and calculable noise properties.
These are all nice features but the
methods explored all presuppose that we have
uniform sampling of the data.  If one has highly inhomogeneous
sampling (and perhaps large holes in the data), one could imagine
modifying the estimators derived here by bending the paths to
avoid holes or noisy regions, but it it not clear that
this is the best thing to do.  A closely related issue is that
we have concentrated on using the shear data on some area to determine
$\kappa$ on the same region, yet (as our aperture mass statistics
clearly show) the shear is also rich in information on $\kappa$
outside the data region.

A quite different approach that can address the above problems
is to attempt to find the most probable solution for the underlying
surface density, given the set of observed shear estimates.
To determine this, we construct the log-likelihood of a model
$\lhood(\kappa) = \log p({\rm data} | {\rm model})$.
An apparent problem is that $\gamma$ is determined in part
by $\kappa$ values outside the grid.  One solution would
be to extend the model to include more grid points, but this
is somewhat costly.  Here instead we exploit the local
expression for $\nabla \kappa$ in terms of $\nabla \gamma_\alpha$
in equation (\ref{eq:gradkappa})

For a $\kappa$ field which is smooth on the scale of the
grid, a grid of estimates of $\hat{\nabla \kappa}$
(obtained {}from the measured shear values by applying
an appropriate discrete difference operator form of the rhs
of equation (\ref{eq:gradkappa})) should be equal to the
same operator applied to the model $\kappa$-grid with
gaussian distributed, but non-trivially correlated,
gaussian residuals.  What we are doing is using the discrete difference
form of equation (\ref{eq:gradkappa}) to project {}from the data that
part which is locally determined, and then solving for the
most likely $\kappa$-configuration assuming we were provided only
with this reduced {}from of the data.
We present two methods to do this.
In \S\ref{sec:MP}, we calculate the most probable $\kappa$ field
assuming a gaussian prior.  This is similar to the method
applied by Kaiser and Stebbins (1991) to reconstruct the local density
field form peculiar velocity data.
We also reformulate in real-space the problem in terms of a standard maximum
likelihood analysis in \S\ref{sec:ML}.

\subsection{Maximum Probability Method}
\label{sec:MP}

We approach the problem {}from the following perspective: given
a set of observations (e.g. shear estimates for $N_{gal}$ galaxies),
we want to {\em fit} a model for the underlying surface density. To
be physically useful, the model must be as unrestrictive and general
as possible. We take
\begin{equation}
\kappa({\bf r}) = \sum a_{\bf k} \cos( {\bf k} \cdot {\bf r} ) +
		b_{\bf k} \sin( {\bf k} \cdot {\bf r} )
		\label{eqn:finite_kappamodel}
\end{equation}
where we assume the surface density is periodic inside some box of
side L -- this is nonrestrictive as we can always zero-pad our data region
to impose periodic boundary conditions.
Allowing $m$ Fourier modes in our model, we rewrite equation
(\ref{eqn:finite_kappamodel}) as
\begin{equation}
\kappa({\bf r}) = \sum_{j=1}^{2m} c_j Y_j({\bf r} ).
\end{equation}
We have complete freedom in determining to Fourier coefficients $c_j$.
A standard approach is to find the most probable set $c_j$ given the
observed data. This implies we minimize
\begin{equation}
\chi^2 = \sum_{i = 1}^{N_{gal}} \frac{1}{\sigma_i^2}
	\biggl[  \gamma_{\alpha i} - \sum_{j=1}^{2m} c_j Y_j({\bf r}_i)
	\chi_{\alpha j }
\biggr]^2
\end{equation}
with respect to $c_k$ so that
\begin{equation}
0 = \sum_{i = 1}^{N_{gal}} \frac{1}{\sigma_i^2}
	\biggl[
	\gamma_{\alpha i} - \sum_{j=1}^{2m} c_j Y_j({\bf r}_i)
	\chi_{\alpha j}\biggr]
	Y_k( {\bf r}_i) \chi_{\alpha k}
\end{equation}
for $\alpha = 0,1$ which labels the two shear components.
This becomes a matrix equation $( A^T A) c = A^T d$ where
$A_{ij} = Y_j( {\bf r}_i ) \chi_{\alpha j} / \sigma_i$ is an
$N_{gal} \times 2m$ matrix and
$d_i = \gamma_{\alpha i}  / \sigma_i$ is a vector of length $2m$. To solve
for the maximum likelihood $c$, we need to invert a $2m \times 2m$ matrix,
which, in general, will be singular for any reasonably large number
of k-modes.

We regularize this inversion with a solution
in the style of maximum entropy reconstructions.
The probability of the model
$c$ given the observed data ${\cal D}$ and some prior information $I$ is
\begin{equation}
P( c  | {\cal D} I ) =
	P( {\cal D} | c I)
	\frac{P( c | I )}{P(  {\cal D  } | I  )}
	\label{eqn:finite_bayest}
\end{equation}
and we want to maximize the probability to determine the
``best fit'' model for $c$. This analysis yields the same equations
as before but we including some prior knowledge to constrain, and
hence regularize, the solutions.
For Gaussian uncertainties of the measurements, the first term on the
rhs of equation (\ref{eqn:finite_bayest}) is as before
\begin{equation}
P( {\cal D} | c I) \propto
	\exp( -\frac{1}{2} \chi^2).
\end{equation}

We take as our prior model $I$ the assumption that
the Fourier coefficients are drawn {}from a Gaussian distribution with
a power spectrum $\langle a_{\bf k}^2 \rangle = \langle b_{\bf k}^2 \rangle =
P_0 k^n$. This adds a diagonal strip to the matrix equation so that
\begin{equation}
( A^T A )_{jk} = \sum_{i = 1}^{N_{gal}}
	\frac{ X_j( {\bf r}_i ) X_k( {\bf r}_i ) }{\sigma_i^2} +
	\frac{\delta_{jk} }{ P(k) }
\end{equation}
and hence the matrix inversion is non-singular for
an arbitrary number of Fourier modes.

This is very similar in spirit to the maximum probability reconstructions
of the density field {}from peculiar velocities employed by Kaiser \&
Stebbins (1991). We adopt a white noise $(n=0)$ power spectrum -- this
is often viewed as a maximally non-committal choice -- and allow the
amplitude to be a varying parameter. Setting large amplitude for the
power (i.e. large $P_0$)
corresponds to little regularization, while a small power amplitude
tends to heavily bias the recovered signal downwards. The optimal amount
to use is determined empirically.
We run extensive simulations for a given $N_{gal}$ to test the sensitivity
of the reconstruction to the amplitude of the power selected. We find
that there is a plateau where the reconstruction is insensitive
to the input power (the reconstruction is stable to variations about this
amplitude). This is a very nice feature as we can tune the regularization
so that the inversion is well behaved, yet we are not forcing the solutions
to some particular form by the choice of our prior.

\subsection{Maximum Likelihood Analysis}
\label{sec:ML}

We now construct an algorithm in real-space to calculate the maximum-likelihood
$\kappa$-field configuration (considered as a grid of $N_x \times
N_y$ values) given a set of shear estimates binned on a similar
grid. We  present the simple
problem where we are given a binned set of shear values with equal noise.
The generalization to unequal noise is straightforward and
is not presented here.
As before, we need to construct the log-likelihood
$\lhood(\kappa) = \log p({\rm data} | {\rm model})$, exploiting
the local relationship between $\grad\kappa$ and observable
shear estimates.

We write the shear-vector estimates as the $2 \times N$ matrix
$\gamma_{\alpha r}$ where $\alpha = 0,1$ labels the shear component
and $r = N_x y + x$ which ranges {}from $0$ to
$N-1$ is a compound index labeling the position
where $N = N_x N_y$ and where the
individual coordinate indices are $x = 0,N_x-1$ and $y = 0, N_y - 1$.
Similarly we denote the model grid of surface density values by the
vector $\kappa_r$.  However, since the discrete differencing
will only determine $\kappa$ modulo an additive constant we
make $\kappa_r$ of dimension $N-1$.  This effectively forces the
corner element $\kappa(x= N_x-1, y=N_y - 1)$ to be zero, though
one is free to adjust the zero-point after the reconstruction to any
chosen reference value.

There are several possible ways to implement the discrete differencing
operator.  We will use
\begin{equation}
\begin{array}{c}
\partial f / \partial x \rightarrow f_{0r} = D_{0rr'} f_r \\
\partial f / \partial y \rightarrow f_{1r} = D_{1rr'} f_r
\end{array}
\end{equation}
where the derivative fields $f_{ir}$ live on an interleaved  $M = M_x \times
M_y$ grid with $M_i = N_i - 1$, and where
\begin{equation}
\begin{array}{c}
D_{0rr'}  = (
(\delta_{x'-x -1, y' -y -1} - \delta_{x'-x, y' -y -1}) +
(\delta_{x'-x -1, y' -y} - \delta_{x'-x, y' -y})
) / 2
\\
D_{1rr'}  = (
(\delta_{x'-x -1, y' -y -1} - \delta_{x'-x -1, y' -y}) +
(\delta_{x'-x, y' -y - 1} - \delta_{x'-x, y' -y})
) / 2
\end{array}
\end{equation}

The discrete difference form of the rhs of equation (\ref{eq:gradkappa}) is
\begin{equation}
\begin{array}{c}
\hat\kappa_{0r} = D_{0rr'} \gamma_{0r'} + D_{1rr'}\gamma_{1r'}\\
\hat\kappa_{1r} = D_{0rr'} \gamma_{1r'} - D_{1rr'}\gamma_{0r'}
\end{array}
\end{equation}
or $\hat\kappa_{ir} = Q_{irj r'} \gamma_{j r'}$
with
\begin{equation}
Q_{i r j r'} =
\begin{array}{c||c|c}
& i = 0 & i = 1 \\
\hline
\hline
j = 0 & D_{0rr'} & -D_{1rr'}\\
\hline
j = 1 & D_{1rr'} & D_{0rr'}\\
\end{array}
\end{equation}
while the lhs is
$\kappa_{ir} = D_{irr'} \kappa_r'$,
with the summation extending {}from $r' = 0$ to $N-2$.

The correlation matrix for the residuals is defined as
\begin{equation}
C_{i r j r'} \equiv \langle
(\hat \kappa_{i r} - \kappa_{i r})
(\hat \kappa_{j r'} - \kappa_{j r'})
\rangle
\end{equation}
and is easily calculated since in the absence of any distorting
influence (and assuming Gaussian distributed independent errors
on the shear-vector estimates)
$\langle \gamma_{lr} \gamma_{mr'} \rangle = \sigma_\gamma^2 \delta_{lm}
\delta_{rr'}$ so
\begin{equation}
C_{i r j r'} = Q_{i r l r''} Q_{j r' m r'''} \langle
\gamma_{lr''} \gamma_{mr'''} \rangle = \delta_{ij} Q_{0rlr''} Q_{0r'lr''}
\end{equation}
and so the log-likelihood is
\begin{equation}
\lhood(\kappa_r) = -{1\over 2} C^{-1}_{r r'}
(Q_{irlr''} \gamma_{lr''} - D_{irr''} \kappa_{r''})
(Q_{ir'mr'''} \gamma_{mr'''} - D_{ir'r'''} \kappa_{r'''})
\end{equation}
where we have defined the $M \times M$ matrix
$C_{rr'} = Q_{0rlr''} Q_{0r'lr''}$.

Minimization of $\lhood$ wrt $\kappa_r$ results in
$A_{r r'} \kappa_{r'} = B_r$
where
$A_{r r'} = D_{ir''r} U_{ir''r'}$ and
$B_r = V_{rlr'} \gamma_{lr'}$
where we have defined
$U_{irr'} = C^{-1}_{rr''} D_{ir''r'}$
and
$V_{rlr'} = Q_{ir''lr'} U_{ir''r}$

Unfortunately, the matrix $A$ is singular; it has
a null space consisting of quasi-oscillatory vectors,
any combination of which dotted with A on left
gives zero on the rhs.
Adding a 9-point Laplacian Tickhonov-Miller regularization
(see Num Recipes, 2nd ed; chapter on inverse methods)
seems to solve this nicely.
We now maximize $\lhood - \alpha L_{rr'} \kappa_{r'} L_{rr''}
\kappa_{r''} / 2$
i.e.~we add to the `chi-squared' an extra term consisting of
some multiple of the
variance in the 2nd derivative field $L_{rr'}\kappa_{r'}$, and
we now obtain the matrix equation
\begin{equation}
\label{eq:Akappa2}
(A + \alpha L^T L) \kappa = B
\end{equation}
where $\alpha$ is the ``trade-off parameter'' and
\begin{equation}	\label{eq:laplacian}
L_{rr'} = 9 \delta(r-r') - \sum\limits_{y=-1}^1\sum\limits_{x=-1}^1
\delta(r + N_x y + x - r')
\end{equation}
is the Laplacian
smoothing operator (we define $L_{rr'}$ for all grid points $r$, but for
for points on the perimeter, where the sum in equation (\ref{eq:laplacian})
would extend outside the grid, we set $L_{r r'} = 0$).

The effect of the regularization is to suppress the spurious
`null-modes' which have strong high frequency components. The trade-off
parameter, $\alpha$, can be thought of as a dial that either makes the solution
very smooth (for large $\alpha$), or allows the solution to fit all
of the data points (for small $\alpha$). For practical implementations,
one wants to choose some intermediate regularization that fits the
data acceptably well, while smoothing out some of the noise
fluctuations and conditioning the inversion process.
In simulations, we find that for $\alpha >\sim 1$ the result is a
heavily smoothed reconstruction.
For small $\alpha$, the inversion is somewhat insensitive to the precise
$\alpha$ value
(unless it is so small that the matrix inversions blow up).
This is an important feature -- the
answers given by this method will not be prejudiced by any reasonable
choice for the regularization, and the user has some freedom in
setting this parameter.

\section{Noise Properties}
\label{sec:noise}

So far we have constructed a whole family of estimators which
all, given perfect data, return an exact reconstruction
of the density field.  The finite-field kernels each measure $\kappa$ relative
to a different baseline, but are all really equivalent since we
can always add any constant we like to force a
particular normalization after the fact. For example, you might
particularly want to measure the surface density in some
region relative to its value in the reference strip shown
in Figure \ref{fig:paths}c. You could of course use the case-c
estimator without further ado. You could equally well use case-b,
measure the value of the reconstruction in your chosen strip and
then readjust the baseline.  If this gives an estimate with
lower statistical uncertainty then you should do so.

Our estimators are weighted sums of observed background ellipticities;
we think of each galaxy giving a noisy estimate of the shear
at its location, the noise arising in part {}from the random intrinsic
ellipticity and in part {}from the measurement error.
The statistical uncertainty depends
on how the weight is distributed among the background
galaxies; generally speaking, the more uniform the
weight the smaller the statistical uncertainty.
Clearly the kernels are always rather sharply peaked towards the
target point.  This generates a $1/r^2$ divergence in the kernels.
This is a common feature for all the methods and is also
largely irrelevant since, as discussed,
it generates predominantly high frequency noise
which goes away when we smooth.  Here we are more interested
in the low spatial frequency behavior of the reconstruction since
this is where the bias appears in equation (\ref{eq:ks93}).
For random background ellipticities the reconstruction will be
a superposition of `inverse kernel' patterns. The low-frequency behavior
of these patterns is somewhat different for the different methods.
In case-c type methods, for a narrow reference strip,
the galaxies in the reference region receive
rather large weight, becoming infinite in the limit
as the width tends to zero. Moreover, the value of the kernel in the annulus
depends only weakly on the target position provided the
target point is not close to the edge.  Thus we would expect that
for a random realization of reference annulus galaxy
ellipticities one will generate a noise fluctuation of random
amplitude, but which will be rather coherent and flat in the center
of the reconstruction.

\subsection{Simulations}

\begin{figure}
\myplotoneb{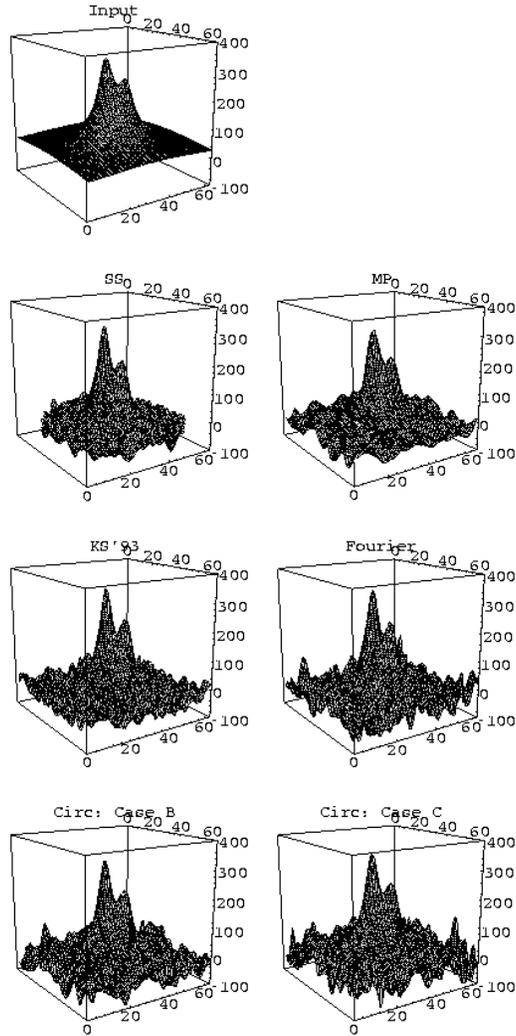}
\caption{Simulation of cluster mass reconstruction.  We have generated
data by laying down randomly placed galaxies and assigning
them ellipticities with a random component to represent the intrinsic
ellipticity plus a systematic component for a model lens.  The reconstruction
is shown for 2500
galaxies at a = 0.9. The lens is a bi-modal softened isothermal like model
at z = 0.2. The main difference among the methods is the noise properties
induced by giving large weight to galaxies near the frame edge.}
\label{fig:simulation}
\end{figure}

As a first, qualitative test we can compare reconstructions
made with simulated data.
To make these simulations we generate random Gaussian
variates $\gamma_1$, $\gamma_2$ for randomly scattered
galaxies.  The rms 1-component shear error
is $\sqrt{\langle \gamma_1^2 \rangle}
\simeq 0.2$. This seems to be quite a good value, determined  observationally
{}from field galaxies.
We then add a small systematic shift calculated
for a particular lens model.  We place the cluster at z~=~0.2 and the
galaxies in a plane at z~=~0.9.
We display the reconstructed surface density using the various methods in
Figure \ref{fig:simulation}. All of the reconstructions have been smoothed
with a 2~pixel Gaussian. The algorithm of Seitz \& Schneider was employed
using circular geometry for calculational ease.

Surprisingly, the bias in
the original KS estimator is not very strong - even though the input
mass distribution is not axisymmetric- and the bias seems only appreciable
in the very corners of the reconstruction.

The results are all somewhat similar in the signal recovery. Of course,
the overall amplitude of the case-c type estimators is higher than for case-b
but this is irrelevant as the baseline surface density is unmeasurable
{}from the shear measurements alone. However, the
the noise properties vary significantly among the methods.
With the case-c estimator, the presence of the
boundary term in the reconstruction gives rise to much higher noise,
especially near the
edge of the reconstruction. The same is true for the Fourier method.
Qualitatively, one favors
the methods that have the smallest
noise fluctuations and the case-c estimators seem
to perform the worst in these simulations.

While these simulations are very simplistic, they are indicative
of the type of behavior we expect.  The various estimation techniques
are all very similar in the signal they recover. However, the case-c
type estimators -- estimators with boundary terms -- tend to have the
most obvious worrisome noise properties.
In the next section, we explore this more quantitatively by calculating
the statistical properties of the noise for the various types of
estimators.

\subsection{Noise Power Analysis} \label{sec:noisepower}

For sufficiently high galaxy number density we expect that
the noise will approach a Gaussian process by virtue of the
central limit theorem.
In that limit everything we need to know about the noise field is
encoded in its power spectrum or autocorrelation function.
As discussed, we are primarily interested in the performance
of these methods for recovering low frequency signals. We do this in two
ways: the first is to calculate the noise-power for the
various methods, and the second is to calculate the variance
in a specific `aperture mass' statistic.

To estimate the noise power we have generated a large number of realizations
for randomly oriented galaxies.  For each of these we estimate the
power and then average the power over the realizations.  In doing this
we came upon a slight technical problem; all of the finite-field methods
tend to
produce `hot-spots' around galaxies close to the edge of the
survey region. These tend to be worse for the case-c type estimator,
but are present in all cases. If we naively measure the power over
the whole survey area we find that these leak though into the low
frequency modes (this is partly due to lack of resolution but is
also partly a real, resolution independent, effect), artificially
enhancing the low frequency power.  This is somewhat discouraging,
since the motivation for developing these methods was to cure the
bias in the KS93 reconstruction, and we seem to have exchanged this
for a lack of precision.

\begin{figure}
\myplotone{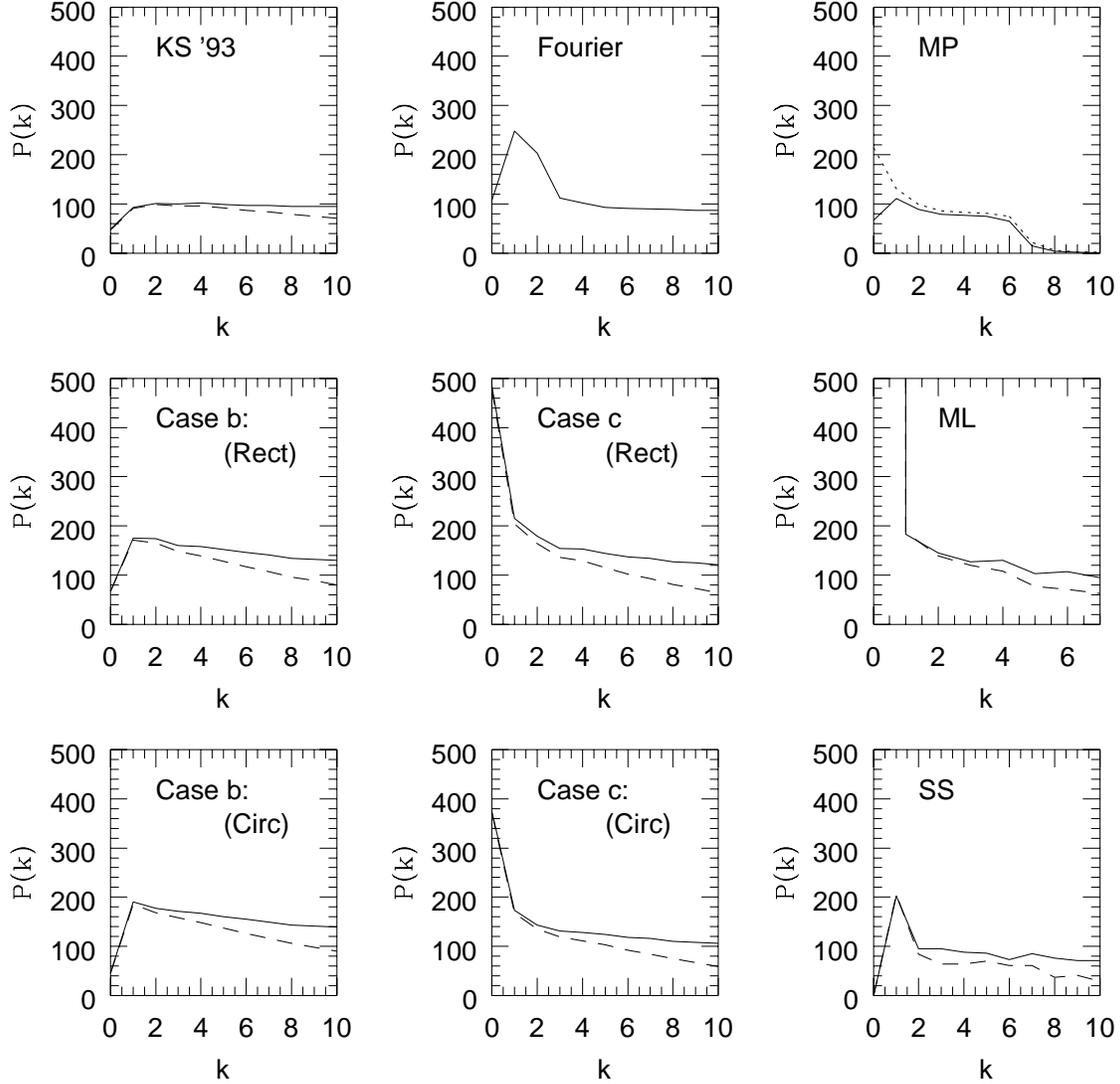}
\caption{The low frequency noise power spectrum for the different
estimators. The plots were made by placing galaxies at random in the
field with an rms polarization of 0.2 per component. The calculations
for each method employed some small scale softening to prevent overflow
in the computer. For example, the case b and c
kernels were softened with $\epsilon = 0.02$ (solid-line) and, $0.04$
(dashed line) but the long wavelength power is unaffected by this
smoothing. Methods with boundary terms in the convolution (case-c, SS)
tend to have larger noise in the zero frequency component.}
\label{fig:power}
\end{figure}

To make a meaningful comparison we have cropped each image to
remove a strip 16 pixels wide around the edge.  The low-frequency
noise power is shown in Figure \ref{fig:power}.
For each finite-field method we have chosen two
different values ($0.02$ and $0.04$) for the softening
parameter $\epsilon$. As expected the low-frequency power is
very insensitive to $\epsilon$ (provided it is small at least).
We have also calculated the noise power using the maximum probability
method with $k_{max} = 10$, and two amplitudes of the power amplitude
that vary by a factor of five. We also display the results using the
maximum likelihood method using trade-off parameters $\alpha=0.001$ and
$\alpha = 0.005$.

Our calculation of the noise power spectrum differs {}from that of
Seitz \& Schneider. In Figure~3 of their paper they make a similar
attempt to calculate the noise power spectrum.
Their results are puzzling however.
For the original KS93
estimator, we expect $P(k) = \; {\rm constant}$ for a pure noise field.
Seitz \& Schneider plot  $k^2 P(k)$, which, judging {}from the points in
their Figure, varies as $\sim k^1$ for long-wavelengths. Indeed
all of the power spectra
they calculate seem to diverge for long wavelengths. One explanation
of this may be the lack of resolution in their simulations.
As we discussed before,
the finite-field kernels have a $1/r^2$ divergence and
a softening of this divergence is required. With coarse resolution, the
$\cos(2 \phi)$ dependence in the kernel is not sampled finely enough
and there is a leakage into the low frequency power. We find that
for all of the methods, it is necessary to use much finer resolution
to compensate for this effect.
Indeed, when we employ the same resolution they used, we reproduce
their curves. With finer resolution however, we obtain somewhat
different results.

The main difference among the noise power in all of the methods is that
the estimators with boundary terms have a much noisier
zero-frequency component.
This is present in the case-c estimators and the Seitz \& Schneider
algorithm.
As discussed, this is because the small number of galaxies
in the annulus give a random fluctuation in the reconstruction
which is nearly independent of target position.  This is rather
interesting.  One motivation for choosing a case-c type estimator
is that for a monotonic cluster profile one would expect this
estimator to give a higher signal.  What we are finding is that the
price for this is increased noise, so this is largely counterproductive.
Similarly, the presence of the boundary term in the Seitz \&
Schneider algorithm introduces a large fluctuation in the long
wavelength noise power.

The maximum likelihood direct reconstruction method has
behavior similar to the other methods at high frequencies. The obvious
difference in this method is the presence of an enormous and very rapid
divergence at small wavenumber.
This suggests that for a pure noise reconstruction, the most likely
solutions determined for the inversion are models with a large
low frequency components.
While a model with, for example, a gradient across
the field seems very unlikely, this seems to be telling us that it is more
likely than any other. This reflects the role of the trade-off
parameter in our regularization -- we can adjust it so that the
reconstruction
is either very smooth (large regularization) or, at the other extreme,
returns a model that fits each and every data point
(and hence has large variations). What we find is that
for a broad choice of trade-off parameters, the result is very stable
and relaxes to a model with long wavelength fluctuations. Dialing down
the trade-off parameter to attempt to reduce the smoothness imposed on
the inversion yields no improvement -- the inversion becomes
ill-conditioned.

The Fourier method has, not surprisingly, a long wavelength fluctuation
as well. This simply comes {}from the fact that we are altering the
data to overcome the periodic boundary condition problem. We have experimented
with cropping the reconstruction area to attempt to remove this
boundary effect and indeed find some improvement. However, there
is some leakage into the long wavelength modes for any reasonable
trimming.

The behavior of the maximum probability method looks most promising. We
see that for reasonable regularization, we obtain a flat noise power
spectrum with the lowest amplitude of all the methods considered here.
It is important to get the regularization correct as we see {}from the
top curve in the maximum probability noise power spectrum -- with too
little amplitude in the prior power, the low frequency modes are amplified.
However, for any given data set with some specified geometry and number
of galaxies, we can determine the required regularization empirically so
that the inversion is well behaved with no bias in the recovered signal.

\section{Aperture Mass Measurement}
\label{sec:apertures}

Two-dimensional images of the mass distribution are nice, but
sometimes one would like to estimate some gross property
such as the mass contained within a given aperture.  Now one could
always take the reconstruction, and integrate over the area in question,
but then calculating the statistical uncertainty
would seem to be a complicated business, particularly if the
reconstruction is smoothed.  In fact
one can construct statistics which measure the mass within an
aperture with a single summation over the galaxies. This is
simpler, and has correspondingly simple statistical properties.
Moreover, these statistics can be designed to use only the data
outside the aperture, which is useful if one wishes to minimize the
contamination of the background galaxy population by faint
cluster members, or to avoid regions where the weak shear
approximation may not be valid.

As discussed in the introduction, {}from equation (\ref{eq:tangentialshear}),
the integral in equation (\ref{eq:ks93}) truncated at inner and outer
radii $r_1$, $r_2$, is equal to $\overline \kappa(r_1) -
\overline \kappa(r_2)$; if we make $r_2$ large this will tend
to $\overline \kappa(r_1)$, and in general gives a lower
bound on $\overline \kappa(r_1)$.
This statistic is quite different in nature
{}from equation (\ref{eq:gradkappa}) which
shows how observations on some region determine the surface
density on the same region (modulo a constant).  Here the
statistic measures, or places a bound on, the mass
in some region with observations taken outside that region.

As we show in appendix \ref{sec:aperturesappendix}, one can derive
equation (\ref{eq:tangentialshear}) {}from Gauss' law. This gives the
mass within an aperture as the integral of the normal component of
$\nabla \phi$ around the aperture boundary, while
the shear measures second derivatives of $\phi$, so it is not
surprising that if we differentiate the mass enclosed with
respect to the aperture radius this picks out some average
of the shear on the boundary.  Here we will first derive equation
(\ref{eq:tangentialshear}) {}from equation
(\ref{eq:kappahat4}) and then generalize it to non-circular apertures.

If we take a circular survey of radius $p$ and place the
origin at the center then $d=p$ and $\gamma^*_\alpha \gamma_\alpha
= \gamma_T$, so
$\gamma^* = \gamma_T$ and equation (\ref{eq:kappahat4})
becomes
\begin{equation}
\kappa(\vec 0) - \overline \kappa(p) = {1\over \pi} \int\limits_0^{p}
dr / r \int d \varphi \gamma_T
\end{equation}
so for concentric disks we find
\begin{equation}	\label{eq:circaperturekappabar}
\overline \kappa(p_1) - \overline \kappa(p_2)
= 2 \int\limits_{p_1}^{p_2} d \ln r \langle \gamma_T \rangle
\end{equation}
where $\langle \ldots \rangle \equiv \int d\varphi / 2 \pi \ldots$ and
which gives a rigorous lower bound on the
mass interior to $r_1$, or, in differential form,
\begin{equation}	\label{eq:dkappabardlnp}
{d \overline \kappa \over d \ln p} = - 2 \langle \gamma_T \rangle
\end{equation}
which is equivalent to equation (\ref{eq:tangentialshear}).

Now $\overline \kappa$ can be thought of as a smoothing of $\kappa$ with
a top-hat window function, and $\overline \kappa(p_1) - \overline \kappa(p_2)$
is a compensated top hat filter.
A general circularly symmetric
smoothing of $\kappa$ can be expressed as
\begin{equation}	\label{eq:generalkappabar}
\int d^2r\; W(r) \kappa =
\int d^2 r\; W'(r) \gamma_T
\end{equation}
with
\begin{equation}	\label{eq:windows}
W'(r) = {2\over r^2} \int\limits_0^r dr'\; r' W(r') - W(r)
\end{equation}
To verify this result, integrate the lhs of equation \ref{eq:generalkappabar}
by parts to obtain an integral involving $\overline \kappa$ and then
integrate by parts once again and replace derivatives of $\overline \kappa$
by $\gamma_T$ using equation (\ref{eq:dkappabardlnp}).

Equation (\ref{eq:windows}) provides a simple way to
construct the KS93-style estimator for
an arbitrarily smoothed reconstruction.
Note that even if $W(r)$ has
compact support $W'$ will still have an extended tail $\propto 1/r^2$ unless
$\int dr\; r W(r)$ vanishes --- i.e.~unless we have a compensated
filter.

The argument leading to equation (\ref{eq:circaperturekappabar}) also
works for non-circular apertures. The key point is that the form
of the kernel in equation (\ref{eq:kappahat4}) depends only on the
{\sl shape\/} of the boundary, so if we consider two
nested boundaries of the same shape and orientation and take the
difference in their $\kappa(\vec 0) - \overline \kappa$ values
there is exact cancellation within the inner boundary curve and
the result will only depend on the data between the curves.
If we let the outer boundary be $\vec p(\varphi)$ and the
inner one be  $\vec p(\varphi)'  = a \vec p(\varphi)$
with scale factor $a < 1$, then
\begin{equation}
\overline \kappa(a) - \overline \kappa(1)
= {1\over A} \int\limits_{r>ap}^{r<p} d^2 r
\gamma^*  p^2 / r^2
\end{equation}
where $A$ is the area inside the boundary $\vec p(\varphi)$,
or in differential form
\begin{equation}
{d \overline \kappa \over d \ln A} = -
\langle \gamma^* \rangle
\end{equation}
where the averaging is understood to be over the thin ribbon
separating two neighboring self-similar curves curves.

It is straightforward to generalize the aperture mass statistic
to inner and outer curves of different shapes; simply choose
some point within the inner aperture as origin, evaluate equation
(\ref{eq:kappahat4}) for the two boundaries and take the difference,
but the resulting statistic will now depend on the data
within the inner aperture.

\section{Discussion}

We have explored the problem of reconstructing the surface density
by direct and regularized inversion methods. The direct methods employ
averaging over line integrals of the shear gradients
exploiting equation (\ref{eq:gradkappa}).  While we have restricted attention
to straight lines, there is still a wide range choice in the
geometry of the reference region.  By performing the integrations
by parts analytically we have been able to show explicitly
how the estimator can be constructed as a weighted sum of
galaxy ellipticities, greatly simplifying the estimation
of the surface density and its uncertainty. We have presented
analytic forms for the kernel for simple rectangular and circular
survey geometries.

Our analysis has revealed a fundamental problem with attempts
to measure $\kappa$ relative to its value on the boundary
of the data region (as in the first method developed by
Schneider).  We find that this is simply not
attainable without introducing very large statistical
uncertainty and/or bias. We have shown however that one can construct
an estimator of $\kappa$ relative to its value in some
finite strip around the boundary, but the performance of this
estimator seems to be no better than a simpler
estimator where the baseline is set to the average over the
whole region (our case-b estimator).  We have also shown how
this particular estimator arises naturally if one develops
(starting {}from Gauss' law) a differential
relation between the mass enclosed within a boundary and the
shear on the boundary.
We have studied a new estimation technique proposed by Seitz \&
Schneider and have shown that it too possesses a boundary term and
suffers {}from the same limitation as the method discussed above.

With fine grid resolution, we calculated the noise power spectrum
for all the of
the reconstruction methods. We find that, in general, estimators such as
the case-c type or the Seitz \& Schneider algorithm which give large
weight to galaxies near the boundary have large long wavelength components
in the noise power. Among the methods that are classified as direct
reconstruction, the case-b type estimators have more desirable noise
properties than any estimator containing the boundary term.

Using a quite different approach, we have shown how one can
construct an exact inverse gradient operator in discrete
Fourier transform space.  This method has some distinct advantages:
it is extremely fast; it can be applied with a very fine grid
to avoid losing any spatial resolution (the real resolution
limit is set by the noisy nature of the shear estimates); and it
is very easily extended to the strong-lensing regime.  Unfortunately,
the low-frequency noise power seems to be somewhat higher than for
the regularized maximum likelihood method for instance.

We also formulated two regularized inversion methods based on the maximum
probability and maximum likelihood techniques.
As with the direct reconstruction methods, we discriminate among the
performance of these estimators by their noise properties.
We find that the maximum likelihood estimator with the usual
9-pt Laplacian regularizer
contains very large low frequency noise components. The maximum probability
method, however, seems to be the most promising. With modest
regularization, the noise power spectrum is flat at all
wavelengths, with essentially no bias in the recovered signal.
Were we given some data and told we could only apply one of the
methods described here then that is what we would choose.  The
method is somewhat more costly in computer time than e.g. KS93
or the Fourier inverse gradient operator, but not unreasonably
so for realistic numbers of galaxies.

We are confident that this paper is not the last word that will
be said on the subject.  The area promises to be rich, simply
because there are potentially infinitely many ways to
recover the surface density {}from the shear. Crudely speaking
this is because we are provided with two real scalar fields
$\gamma_1$ and $\gamma_2$ which is much more information than
we need to recover the single scalar $\kappa$, so there are
many ways one can squander this information and yet still
recover an unbiased answer.  All we have done here is to
apply a geometric construction to generate a family of viable
estimators, and we have then compared their statistical
performance.  We have no way of knowing whether there might
not be much better estimators out there somewhere, but perhaps
the similarity in noise power for the methods we have
considered is telling us something.

\acknowledgments{We would like to thank P. Schneider, C. Seitz and S. Seitz
for enlightening and helpful conversations.}

\appendix

\section{Aperture Masses {}from Gauss' Law}
\label{sec:aperturesappendix}

Consider a circular loop of radius $r$ and with azimuthal
coordinate $\varphi$.  At each point on the circle construct
locally cartesian coordinates $n,t$ along the outward
normal and tangential directions $\hat n, \hat t$.  By Gauss'
law, the mass contained within the loop is
\begin{equation}
M(r) = \int^r d^2 r\; \kappa(\vec r) = {r\over 2} \oint d\varphi \phi_{,n}
\end{equation}
so
\begin{equation}
{dM\over dr} = {M\over r} + {r\over 2} \oint d\varphi \phi_{,nn}
\end{equation}
but $\phi_{,nn} = \kappa - \gamma_T$ where the {\sl tangential
shear\/} is $\gamma_T = (\phi_{,tt} - \phi_{,nn}) / 2$.  In components
in the general coordinate frame $\gamma_T = -(\gamma_1 \cos 2 \varphi
+ \gamma_2 \sin 2 \varphi$). Hence
\begin{equation}
\label{eq:tangentialshear2}
\langle \gamma_T \rangle
= {-1\over 2 \pi r}({dM\over dr} - {2M\over r}) = {-1\over 2} {d \overline
\kappa
\over d \ln r}
\end{equation}
where $\langle \gamma_T \rangle =  \int d \varphi \gamma_T / 2 \pi$ and
$\overline \kappa \equiv \int d^2 r \kappa / \int d^2 r$.
This is equivalent to equation (\ref{eq:tangentialshear}).

Let us now generalize this to non-circular apertures.
Take some arbitrary point as origin and consider the self-similar
family of closed curves $\vec r(a,\lambda) = a \vec c(\lambda)$ where
$\lambda$ is a cyclic parameter around the curve and $a$ is a scale
factor. If we consider two neighboring curves then vector
connecting points with the same $\lambda$ has length $\delta r =
c(\lambda) \delta a$, and if we erect orthogonal locally
normal and tangential coordinates $(n,l)$, such pairs have
separation $\delta n = c \delta a \sin \psi$, $\delta l = \delta n
\cot \psi$ where $\cos \psi = \vec c \cdot \vec c' / c c'$
and where $\vec c' = d\vec c / d \lambda$ and $c'= |\vec c'|$.
By Gauss' law,
\begin{equation}
M(a) = {a\over 2} \oint d \lambda c' \phi_{,n}(\vec r)
\end{equation}
and
\begin{equation}
M(a') = {a'\over 2} \oint d \lambda c' \phi_{,n}(\vec r +
\delta n \hat n + \delta l \hat l)
\end{equation}
where $\hat n, \hat l$ are unit vectors.
Taylor expanding $\phi_{,n}$ to 1st order and replacing $\phi_{,nn}$ by
$\kappa - \gamma_+$ with parallel component of the shear
$\gamma_+$ defined to be $(\phi_{,ll} - \phi_{,nn})/2$, and
with the orthogonal component of the shear
$\gamma_\times = - \phi_{,nl}$, we
have
\begin{equation}
\delta M = M(a) \delta a / a + {1\over 2} \int d\lambda a c' \delta n
(\kappa - \gamma_+ - \gamma_\times \cot \psi)
\end{equation}
but the integral here is just the integral of
$\kappa - \gamma_+ - \gamma_\times \cot \psi$ over the area $\delta A$
in the ribbon between the two curves, and we therefore have,
for $\delta \overline \kappa = \delta(M/A)$,
\begin{equation}
\delta \overline \kappa = {-1\over A} \int_{\delta A} d^2 r (
\gamma_+ + \gamma_\times \cot \psi)
\end{equation}
or
\begin{equation}	\label{eq:dkappadlnA}
{d \overline \kappa \over d \ln A} =
\langle \gamma_+ + \gamma_\times \cot \psi \rangle
\end{equation}
where the averaging is understood to be over the thin ribbon.

Now the area is $A = A_0 (r^2 / c^2)$ where $A_0$ is the area enclosed
within the curve $\vec r = \vec c$, so we have
\begin{equation}
\overline \kappa(a) - \overline \kappa(1)
= {1\over A_0} \int\limits_{r>ac}^{r<c} d^2 r (
\gamma_+ + \gamma_\times \cot \psi) c^2 / r^2
\end{equation}
If we take $\vec c$ to be the perimeter of the data region
and let $a \rightarrow 0$ then we obtain a measure
of $\kappa$ at a point (the origin) relative to
the mean over the data region and the result is precisely equivalent to
equation (\ref{eq:kappahat4}).
For finite $a$ we obtain  the mass enclosed within the
aperture $a \vec c(\lambda)$ as an integral over the data
lying outside the aperture just as in \S\ref{sec:apertures}.

The alternative derivation
given in this section is not entirely pointless, however, since because of the
redundancy in the data --- we appear to have a two component field
$\gamma_\alpha(\vec r)$ but in reality both $\gamma_1$ and $\gamma_2$
are derived {}from a single
scalar surface potential function $\phi(\vec r)$ ---
there are many different estimators which measure the
same physical quantity but which have different noise characteristics.
A simple example of this is equation (\ref{eq:tangentialshear})
to which we could, if we were perverse,
add any multiple of the integral of $\gamma_\times$ around the loop
which physically must vanish and so, with real data, will add pure noise.
It was therefore not entirely
a foregone conclusion that these two different approaches would
give the same result, but in fact they do.

\end{document}